%% file: main.tex
\documentclass{article}
\usepackage{spconf,amsmath,graphicx}
\usepackage{amsmath,graphicx}
\usepackage{amssymb}
\usepackage{hyperref}
\usepackage{cite}
\usepackage{url}
\usepackage{caption}
\usepackage{subcaption}
\usepackage{microtype}
\usepackage{enumitem}
\usepackage{booktabs}
\usepackage{tabularx}
\usepackage{multirow}

\newcolumntype{C}{>{\centering\arraybackslash}X}
\newcolumntype{L}{>{\raggedright\arraybackslash}X}
\newcolumntype{R}{>{\raggedleft\arraybackslash}X}
\newcolumntype{P}[1]{>{\raggedright\arraybackslash}p{#1}}
\usepackage{siunitx}
\usepackage{etoolbox}                           
\newcommand{\ubold}{\fontseries{b}\selectfont}  
\robustify\ubold                                

\newcommand{\tablecaptionsep}{\vspace*{-5pt}}

\usepackage{xcolor}
\definecolor{hermancolor}{HTML}{FF6600}

\definecolor{mattcolor}{HTML}{0004ff}

\definecolor{shorten}{HTML}{7a7a7a}

\mathchardef\mhyphen="2D

\def\modelname{{ASGAN}}

\title{GAN you hear me? \\ Reclaiming unconditional speech synthesis from diffusion models}

\name{Matthew Baas and Herman Kamper\thanks{All experiments were performed on Stellenbosch University's High Performance Computing (HPC) cluster.}}

\address{MediaLab, Electrical \& Electronic Engineering, Stellenbosch University, South Africa}

\copyrightnotice{978-1-6654-7189-3/22/\$31.00~\copyright2023 IEEE}

\begin{document}
\ninept
\maketitle
\input{0_abstract}

\section{Introduction}
\label{sec:intro}

\input{1_introduction}

\section{Related work}
\label{sec:related_work}

\input{2_related_work}

\vspace{-1.3mm}
\section{\modelname{}: Audio Style GAN}
\label{sec:model}
\vspace{-0.7mm}

\input{3_model}

\vspace{-1.5mm}
\section{Experimental Setup}
\label{sec:exp_setup}
\vspace{-1mm}

\input{4_exp_setup}

\vspace{-1.5mm}
\section{Results: Unconditional speech synthesis}
\label{sec:5_results}

\input{5_results}

\vspace{-1.3mm}
\section{Unseen tasks: \\voice conversion and speech editing}\label{sec:6_unseen}

\input{6_zeroshot}

\vspace{-1.7mm}
\section{Conclusion}\label{sec:7_conclusion}

\input{7_conclusion}

\newpage

\bibliographystyle{IEEEbib}
\bibliography{references}

\end{document}

%% file: 0_abstract.tex
\begin{abstract}
We propose AudioStyleGAN (ASGAN), a new generative adversarial network (GAN) for unconditional speech synthesis.
As in the StyleGAN family of image synthesis models, ASGAN maps sampled noise to a disentangled latent vector which is then mapped to a sequence of audio features so that signal aliasing is suppressed at every layer.
To successfully train ASGAN, we introduce a number of new techniques, including a modification to adaptive discriminator augmentation to probabilistically skip discriminator updates.
ASGAN achieves state-of-the-art results in unconditional speech synthesis on the Google Speech Commands dataset.
It is also substantially faster than the top-performing diffusion models.
Through a design that encourages disentanglement, ASGAN is able to perform voice conversion and speech editing without being explicitly trained to do so.
ASGAN demonstrates that GANs are still highly competitive with diffusion models.
Code, models, samples: {\footnotesize \url{https://github.com/RF5/simple-asgan/}}.
\end{abstract}

%% file: 1_introduction.tex
Unconditional speech synthesis is the task of generating coherent speech without any conditioning inputs such as text or speaker labels~\cite{adv_audio_synth_donahue2018adversarial}.
As
in 
image synthesis \cite{fid_heusel2017gans}, a well-performing unconditional speech synthesis model would have several useful applications:
from latent interpolations between utterances and fine-grained tuning of different aspects of the generated speech, to audio compression and better
probability density estimation of speech. 

Spurred on by recent improvements in diffusion models \cite{diffusion_sohl2015deep} for images \cite{dall-e_ramesh2021zero,dall-e-2_ramesh2022hierarchical,imagen_saharia2022photorealistic}, there has been a substantial improvement in 
unconditional speech synthesis 
in the last few years.
The
current best-performing approaches are all trained as
diffusion models \cite{sashimi_goel2022s, diffwave_kong2020}.
Before this, most studies
used generative adversarial networks (GANs) \cite{gans_goodfellow2014generative} 
that map
a latent vector to a sequence of speech features with a single forward pass through the model.
However,
performance was limited \cite{adv_audio_synth_donahue2018adversarial, beguvs2020generative}, leading to GANs falling out of favour for this task. 

Motivated by the StyleGAN literature \cite{stylegan1_karras2019style,stylegan2_karras2020analyzing,stylegan3_karras2021alias} for image synthesis, we aim to reinvigorate GANs for unconditional speech synthesis.
To this end, we propose AudioStyleGAN (\modelname{}): a convolutional GAN which maps a single latent vector to a sequence of audio features, 
and is designed to have a 
disentangled latent space.
The model is based in large part on StyleGAN3 \cite{stylegan3_karras2021alias}, which we adapt for audio synthesis.
Concretely, we
adapt 
the style layers to
remove signal 
aliasing caused by the non-linearities in the network.
This is accomplished with anti-aliasing filters to ensure that the Nyquist$\mhyphen$Shannon sampling limits are met in each layer.
We also propose a modification to adaptive discriminator augmentation \cite{ada_karras2020training} to stabilize training by randomly dropping discriminator updates based on a guiding signal.

Using objective metrics to measure the quality and diversity of generated samples~\cite{is_salimans2016improved,fid_heusel2017gans,am_zhou2018activation}, we show that ASGAN sets a new state-of-the-art in unconditional speech synthesis on 
the Google Speech Commands digits dataset~\cite{speechcommands_warden2018speech}.
It not only outperforms the best existing models, but is also faster to train and faster in inference.
Mean opinion scores (MOS) 
also indicate
that \modelname{}'s generated
utterances sound
more natural (MOS: 3.68) than the existing best model (SaShiMi~\cite{sashimi_goel2022s}, MOS: 3.33).

Through \modelname{}'s design, the model's latent space is disentangled during training, enabling the model -- without any additional training 
-- to also perform voice conversion and
speech editing
in a zero-shot fashion.
Objective metrics that measure latent space disentanglement indicate that \modelname{} has smoother latent representations 
compared to
existing diffusion models.

%% file: 2_related_work.tex
\begin{figure*}[t!]
\centering
\centerline{\includegraphics[width=0.999\linewidth]{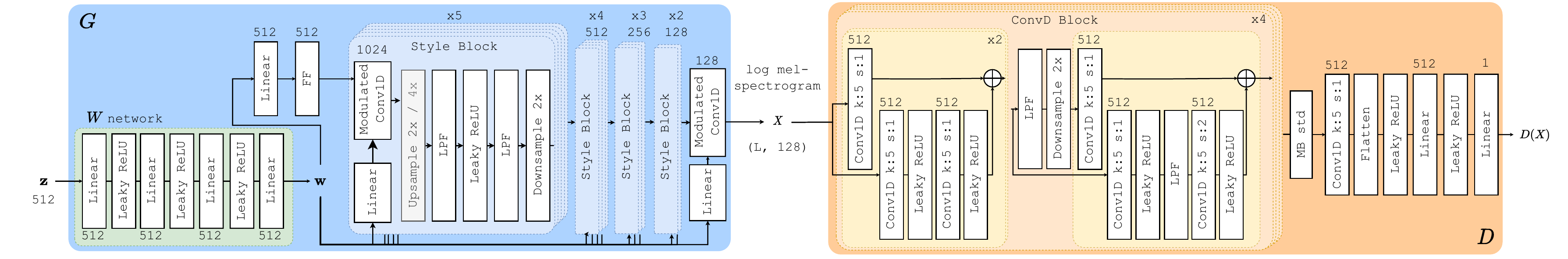}}
\caption{
    The \modelname{} generator (left) and discriminator (right). 
    FF, LPF, Conv1D indicate Fourier feature \cite{stylegan3_karras2021alias}, low-pass filter, and 1D convolution layers, respectively.
    The numbers above linear and convolutional layers indicate the number of output features/channels for that layer.
    Stacked blocks indicate a layer repeated sequentially, with the number of repeats indicated above the block (e.g. ``\texttt{x3}").
}
\label{fig:1_model_arch}
\vspace{-4mm}
\end{figure*}

We start by distinguishing what we call \textit{unconditional speech synthesis} to the related but different task of \textit{generative spoken language modeling} (GSLM).
In GSLM, a large autoregressive language model is typically trained on some discrete units
(e.g.\ HuBERT~\cite{hubert2021} clusters or clustered spectrogram features),
similar to how a language model is trained on text~\cite{textless_nlp2_polyak2021speech,textless_nlp3_kharitonov-etal-2022-text}.
While this also enables the generation of speech without any conditioning input, GSLM implies a 
model structure consisting of an encoder to discretize speech,
a language model, and a decoder \cite{textless_nlp1_lakhotia2021generative}.
This means that during generation, you are bound by the discrete units in the model.
E.g.,
it is not possible to interpolate between two utterances in a latent space
or to directly
control speaker characteristics during generation.
If this is desired, additional components must be explicitly built into the model~\cite{textless_nlp3_kharitonov-etal-2022-text}.

In contrast, in unconditional speech synthesis we 
do not assume any knowledge of particular aspects of speech beforehand. 
Instead of using some intermediate discretization step, such models typically
use noise to directly generate speech, 
often via some latent representation.
The latent space should ideally be disentangled, 
allowing for modelling and control of the generated speech.
In contrast to GSLM, the synthesis model should learn to disentangle without being explicitly designed 
to control specific speech characteristics.
In some sense this is a more challenging task than GSLM, which is why most unconditional speech synthesis models are still evaluated on short utterances of isolated spoken words~\cite{adv_audio_synth_donahue2018adversarial} (as we also do here).

Within 
unconditional speech synthesis,
a substantial body of work focuses on either autoregressive \cite{oord2016wavenet} models -- generating a current sample based on previous outputs --
or diffusion models~\cite{diffwave_kong2020}. 
Diffusion models iteratively de-noise a 
sampled signal into a
waveform through a Markov chain with a constant number of steps~\cite{diffusion_sohl2015deep}.
At each inference step, the original noise signal is slightly de-noised until -- in the last step -- it resembles coherent speech.
Autoregressive and diffusion
models
are 
relatively slow
because they require
repeated forward passes through the
model during inference.

Earlier studies~\cite{adv_audio_synth_donahue2018adversarial,beguvs2020generative}
attempted
to use 
GANs \cite{gans_goodfellow2014generative} for unconditional speech synthesis, which has the advantage of requiring only a single pass through the model.
While 
results showed some initial promise, performance was poor
in terms of speech quality and
diversity, with the more recent diffusion models performing much better~\cite{sashimi_goel2022s}.
However, there have been substantial improvements in GAN-based modelling for image synthesis
in the intervening years~\cite{ada_karras2020training, stylegan1_karras2019style, stylegan2_karras2020analyzing}.
Our goal is to
improve the performance of the earlier GAN-based unconditional speech synthesis models by adapting 
lessons from these recent image synthesis studies.

Some of these 
innovations in GANs
are
modality-agnostic:
$R_1$ regularization \cite{r1_reg_mescheder2018training} and 
exponential moving
averaging
of generator weights \cite{ema_gen_weights_karras2018progressive} can be directly transferred from the vision domain to speech.
Other techniques, such as the 
carefully designed anti-aliasing filters between layers in StyleGAN3 \cite{stylegan3_karras2021alias} 
require specific adaptation;
in contrast to images, there is little meaningful information in speech below 300~Hz, 
necessitating a redesign of the anti-aliasing filters.

In a very related research direction, Beguš~\cite{BEGUS2022101244, beguvs2020generative} has been studying how GAN-based unconditional speech synthesis models internally perform lexical and phonological learning, and how this relates to human learning.
These studies, however, have been relying on the older GAN synthesis models.
We hope that by developing better performing GANs for unconditional speech synthesis, such investigations
will also be improved.
Recently, \cite{audio_stylegan2_2022} attempted to directly use StyleGAN2 for conditional and 
unconditional synthesis of emotional vocal bursts.
This further motivates a reinvestigation of GANs, but here we look specifically at the generation of speech rather than paralinguistic sounds.

%% file: 3_model.tex
Our model is based on the StyleGAN 
family of models \cite{stylegan1_karras2019style} for image synthesis. 
We adapt and extend the approach to audio, and therefore dub our model
AudioStyleGAN (\modelname{}).
The model follows the setup of a standard GAN with a single generator network $G$ and a single discriminator $D$ \cite{gans_goodfellow2014generative}.
The generator $G$ accepts a vector $\mathbf{z}$ sampled from a normal distribution and processes it into a sequence of speech features $X$.
In this work, we restrict the sequence of speech features $X$ to always have a fixed pre-specified duration.
The discriminator $D$ accepts a sequence of speech features $X$ and yields a scalar output.
$D$ is optimized to raise its output for $X$ sampled from real data and lower its output for $X$ produced by the generator.
Meanwhile, $G$ is optimized to maximize $D(X)$ for $X$ sampled from the generator, i.e.\ when $X=G(\mathbf{z})$.
The features $X$ are converted to a waveform using a pretrained HiFi-GAN vocoder~\cite{hifi-gan}.
During training, a new adaptive discriminator updating technique is added to ensure stability and convergence, as 
discussed in Sec.~\ref{sec:exp_setup}.

\subsection{Generator}

The architecture of the generator $G$ is shown on the left of Fig.~\ref{fig:1_model_arch}.
It consists of a latent mapping network $W$ that converts $\mathbf{z}$ to a disentangled latent space, a special Fourier feature (FF) layer which converts a single vector from this latent space into a sequence of cosine features of fixed length, and finally a convolutional encoder which iteratively refines the cosine features into the final speech features $X$.

\textbf{Mapping network:}
The mapping network $W$ is a 
multi-layer perceptron
with leaky ReLU activations.
As input it takes in a vector sampled from a normal distribution 
$\mathbf{z} \sim Z = \mathcal{N}(\mathbf{0}, \mathbf{I})$; 
we use a 512-dimensional multi-variate normal vector, $\mathbf{z} \in \mathbb{R}^{512}$.
Passing $\mathbf{z}$ 
through the mapping network produces a latent vector $\mathbf{w} = W(\mathbf{z})$ of the same dimensionality as $\mathbf{z}$.
As explained in~\cite{stylegan1_karras2019style}, 
the primary purpose of $W$
is to learn to 
map noise to a linearly disentangled 
space,
as 
this will allow for 
controllable and understandable synthesis.
$W$ is coaxed into learning such a disentangled representation because it can only \textit{linearly} modulate channels of the cosine features in each layer of the convolutional encoder (see 
details below).
This means
that $W$ 
must learn
to map
the random normal $Z$-space into a $W$-space that 
linearly disentangles common factors of speech variation.

\textbf{Convolutional encoder:}\label{sec:3_conv_encoder}
The convolutional encoder begins by linearly projecting $\mathbf{w}$ as the input to an FF layer \cite{fourierfeat_tancik2020}. 
We use the Gaussian Fourier feature mapping~\cite{fourierfeat_tancik2020} and incorporate the transformation from 
StyleGAN3 \cite{stylegan3_karras2021alias}.
The Gaussian FF layer samples a frequency and phase from a Gaussian distribution for each output channel. 
The layer then linearly projects the input vector to a vector of phases which are added to the random phases.
The output is calculated as the cosine functions of these frequencies and phases, one frequency/phase for each output channel.
The result is that
$\mathbf{w}$
is converted
into a 
sequence of vectors at the output of the FF layer.
This 
is 
iteratively passed through several \texttt{Style Blocks}.
In each \texttt{Style Block} layer, the input sequence 
is
passed through a modulated convolution layer \cite{stylegan2_karras2020analyzing} whereby 
the final convolution kernel is computed by multiplying the layer's learnt kernel with the style vector derived from $\mathbf{w}$, broadcasted over the length of the kernel.
To ensure the signal does not experience aliasing 
due to
the non-linearity, the leaky 
ReLU layers are surrounded by layers responsible for anti-aliasing (explained below).
All these layers comprise a \texttt{Style Block}, which is repeated in groups of 5, 4, 3, and finally 2 blocks.
The last block in each group upsamples by $4\times$ instead of $2\times$, thereby increasing the sequence length by a factor of 2 for each group.
A
final 1D convolution 
projects the output from the last group
into the audio feature space (e.g. log mel-spectrogram
or 
HuBERT features \cite{hubert2021}), as illustrated in the middle of Fig.~\ref{fig:1_model_arch}.

\textbf{Anti-aliasing filters:}
From 
image synthesis with GANs 
\cite{stylegan3_karras2021alias}, we know that the generator 
must include anti-aliasing filters 
for the signal propagating through the network to 
satisfy the Nyquist-Shannon sampling theorem.
This is why, before and after a 
non-linearity, we include upsampling, low-pass filter (LPF), and downsampling layers in each \texttt{Style Block}.
The motivation from 
\cite{stylegan3_karras2021alias}
is that most non-linearities 
introduce arbitrarily
high-frequency information into the output signal.
The signal we are modelling (speech) is
continuous,
and the internal discrete-time features 
that are passed through the network is therefore a digital representation of this continuous signal.
From the Nyquist-Shannon sampling theorem, we know that for such a discrete-time signal to accurately reconstruct the continuous signal, it must be bandlimited to $\SI{0.5}{\text{cycles/sample}}$.
If not, the generator can learn
to use 
aliasing artifacts
to fool the discriminator, to the detriment of the quality and control of the final output. 
To address this, we follow 
\cite{stylegan3_karras2021alias}: we approximate an ideal continuous LPF through the sequence of upsample, LPF, non-linearity, and downsample operations to ensure that the signal is bandlimited.
We reason that the generator should ideally first focus on generating course features before generating good high-frequency details, which will inevitably contain more trace aliasing artifacts.
So 
we design the filter cutoff 
to begin at a small value in the first \texttt{Style Block}, and increase gradually to near the critical Nyquist frequency in the final block.

\vspace{-2.2mm}
\subsection{Discriminator}

The discriminator $D$ has 
a convolutional architecture similar to 
\cite{stylegan2_karras2020analyzing}, taking a sequence of speech features $X$ as input and predicting whether it is generated by $G$ or sampled from the dataset.
Concretely, $D$ consists of four \texttt{ConvD Blocks} and a network head, as show in
Fig.~\ref{fig:1_model_arch}.
Each \texttt{ConvD Block} is comprised of 1D convolutions with skip connections, and a downsampling layer with an anti-aliasing LPF in the last skip connection.
The LPF cutoff 
is set as the Nyquist frequency for all layers.
The
number of layers and channels
are chosen so that $D$ has roughly the same number of parameters as $G$.
$D$'s
head consists of a minibatch standard deviation \cite{ema_gen_weights_karras2018progressive} layer and a 1D convolution layer before passing the flattened activations through a final linear projection head to arrive at the logits.
Both $D$ and $G$ are
trained using the non-saturating logistic loss \cite{gans_goodfellow2014generative}.

\vspace{-1.5mm}
\subsection{Vocoder}

Once the generator $G$ and discriminator $D$ are trained, we need a way to convert the 
speech features back to waveforms.
For this
we use a pretrained HiFi-GAN vocoder \cite{hifi-gan} that vocodes either log mel-scale spectrograms or HuBERT features \cite{hubert2021}.

%% file: 4_exp_setup.tex
\subsection{Data}

To compare to existing unconditional speech synthesis models, we use the Google Speech Commands dataset of isolated spoken words \cite{speechcommands_warden2018speech}.
As in other studies \cite{sashimi_goel2022s, diffwave_kong2020, adv_audio_synth_donahue2018adversarial}, we use the
subset corresponding to the ten spoken digits ``zero'' to ``nine'' (called SC09).
The digits are
spoken by various speakers 
under different channel conditions.
This makes it a challenging benchmark for unconditional speech synthesis.
All utterances are roughly a second long and are sampled at 16~kHz.

\vspace{-1.2mm}
\subsection{Evaluation metrics}\label{sec:4.2_eval_metrics}

We train and validate our models on the official training split from SC09. We then evaluate 
speech synthesis quality by seeing how well newly generated utterances match the distribution of the SC09 test split.
We use metrics
similar to those for image synthesis; 
they 
measure either the \textit{quality} of generated utterances (realism compared to test data), or the \textit{diversity} of generated utterances (how varied the 
utterances are relative to the test set),
or a combination of both.

These 
metrics 
require extracting features or predictions from a supervised speech classifier network trained to classify the utterances from SC09.
While there is no consistent pretrained classifier used for this purpose, 
we opt
to use a 
ResNeXT architecture \cite{resnext_xie2017aggregated}, similar to previous studies \cite{sashimi_goel2022s, diffwave_kong2020}.
The trained model has 
a 98.1\% word classification accuracy on the SC09 test set, and we make the model available for future comparisons.\footnote{\scriptsize\url{https://github.com/RF5/simple-speech-commands}} 
Using either the classification output or 1024-dimensional features extracted from the penultimate layer in the classifier,
we consider the following metrics.

\textit{Inception score} (IS) measures the diversity and quality of generated samples by evaluating the Kullback-Leibler (KL) divergence between the label distribution from the classifier output and the mean label distribution over a set of generated utterances \cite{is_salimans2016improved}.
\textit{Modified Inception score} (mIS) extends IS by incorporating a measure of intra-class diversity (in our case over the ten digits) to reward models with a higher intra-class entropy \cite{modified_is_gurumurthy2017deligan}.
\textit{Fréchet Inception distance} (FID) computes a measure of how well the distribution of generated utterances matches the train-set utterances by comparing the classifier features of generated and real data \cite{fid_heusel2017gans}.
\textit{Activation maximization} (AM) measures generator quality by comparing the KL divergence between the classifier class probabilities from real and generated data, while penalizing high classifier entropy samples produced by the generator \cite{am_zhou2018activation}. 
Intuitively, this attempts to account for class imbalance in the training set and also intra-class diversity.
All these metrics have been used in previous unconditional speech synthesis studies~\cite{sashimi_goel2022s, diffwave_kong2020}.

A major 
motivation for 
\modelname{}'s design is 
latent-space disentanglement.
To evaluate this,
we 
use
two disentanglement metrics on the latent
$Z$-space and $W$-space.
\textit{Path length}
measures the mean $L_2$ distance moved by the classifier features when the latent point ($\mathbf{z}$ or $\mathbf{w}$) 
is randomly perturbed
slightly, averaged over many 
perturbations \cite{stylegan1_karras2019style}.
A lower value
indicates
a smoother latent space.
\textit{Linear separability} 
utilizes a linear support vector machine (SVM) 
to classify the digit of a latent point.
The metric is computed as the additional information (in terms of mean entropy) necessary to correctly classify an utterance (in terms of which digit is spoken) \cite{stylegan1_karras2019style}.
Again, a lower value indicates a more linearly disentangled latent space.
These metrics are averaged over 5000 generated utterances for each model.
As in \cite{stylegan1_karras2019style},
for linear separability we exclude half the generated utterances for which the ResNeXT classifier is least confident.

Finally, to give an indication of naturalness,
we compute an estimated mean opinion score (eMOS) using a pretrained \texttt{Wav2Vec2 small} baseline
from the recent VoiceMOS challenge~\cite{huang2022voicemos}.
This model is trained to predict the naturalness score that a human would assign to an utterance from
1 (least natural) to 5 (most natural).
We also perform a subjective MOS evaluation using the same scale.
Concretely, we utilize Amazon Mechanical Turk to obtain 
240 opinion scores for each model with 12 speakers listening to each utterance.

\vspace{-1mm}
\subsection{Baselines}

We compare to the following unconditional speech synthesis methods (Sec.~\ref{sec:related_work}):
WaveGAN \cite{adv_audio_synth_donahue2018adversarial}, DiffWave \cite{diffwave_kong2020}, autoregressive SaShiMi and Sashimi+DiffWave~\cite{sashimi_goel2022s} (the current best performing model on SC09).
For WaveGAN we use 
the trained model provided by the 
authors \cite{adv_audio_synth_donahue2018adversarial}, while for DiffWave we use an 
open-source pretrained 
model.\footnote{\scriptsize
\url{https://github.com/RF5/DiffWave-unconditional}
}
For the autoregressive SaShiMi model, we use the 
code provided by the authors to train an unconditional SaShiMi model on SC09 
for 1.1M updates \cite{sashimi_goel2022s}.\footnote{\scriptsize
\url{https://github.com/RF5/simple-sashimi}
}
Finally, for 
SaShiMi+DiffWave, 
we modify the autoregressive SaShiMi code
and
combine it with DiffWave 
according to \cite{sashimi_goel2022s}; we train
it on SC09 for 800k updates with the parameters in the original paper \cite{sashimi_goel2022s}.\footnotemark[3]
In all experiments, we perform direct sampling from the latent space for the GAN and diffusion models according to the original papers.
For the autoregressive models, we directly sample from the predicted output distribution for each time-step sample.

\vspace{-1mm}
\subsection{\modelname{} implementation}

We train two variants of our model: a log mel-spectrogram based model and a HuBERT feature based model \cite{hubert2021}.
The former
is shown in Fig.~\ref{fig:1_model_arch}, where
the model outputs
128 mel-frequency bins at 
a hop and window size of 
10~ms and 64~ms, respectively.
The HuBERT 
model is identical
except 
that it only uses
half the sequence length
(since HuBERT features are 20~ms instead of the 10~ms spectrogram frames) 
and 
has
a different number 
of output channels in the four groups of \texttt{Style Block}s: $[1024, 768, 512, 512]$ convolution channels instead of $[1024, 512, 256, 128]$.

The HiFi-GAN vocoder for both the HuBERT and mel-spectrogram
model is
based on the original 
implementation \cite{hifi-gan}.
The HuBERT HiFi-GAN is trained on the Librispeech \texttt{train-clean-100} subset~\cite{panayotov2015librispeech}
to vocode activations from layer 6 of the 
HuBERT \texttt{Base} model \cite{hubert2021}.
The mel-spectrogram HiFi-GAN is trained on the Google Speech Commands dataset.

Both 
\modelname{} variants are trained with Adam \cite{adam} ($\beta_1 = 0, \beta_2 = 0.99$), clipping gradient norms at 10, and a learning rate of $3\cdot10^{-3}$ for 520k iterations with a batch size of 32. 
Several 
tricks are
used to stabilize GAN training:
(i) equalized learning rate
\cite{ema_gen_weights_karras2018progressive}, (ii) leaky ReLU activations with $\alpha=0.1$, (iii) exponential moving averaging 
for the generator weights \cite{ema_gen_weights_karras2018progressive}, (iv) $R_1$ regularization \cite{r1_reg_mescheder2018training}, and (v)
a 0.01-times smaller learning rate for
the mapping network $W$
\cite{stylegan3_karras2021alias}.

We also introduce a new technique for updating the discriminator.
Concretely, we first scale $D$'s learning rate by 0.1 compared to the generator as otherwise we find it overwhelms 
$G$
early on in training.
Additionally we employ a dynamic method for updating $D$, inspired by adaptive discriminator augmentation \cite{ada_karras2020training}:
during each iteration, we skip $D$'s update with probability $p$.
The probability $p$ is initialized at 0.1 and is updated every 16th generator step or whenever the discriminator is updated.
We keep a running average $r_t$ of the proportion of $D$'s outputs on real data $D(X)$ that are \textit{positive} (i.e. that $D$ can confidently identify as real).
Then, if $r_t$ is greater than 0.6 we increment $p$ by 0.05 (capped at 1.0), and if $r_t$ is less than 0.6 we decrease $p$ by 0.05 (capped at 0.0).
In this way we \textit{adaptively skip discriminator updates}. When $D$ becomes too strong, $r_t$ and $p$ rise, and so $D$ is updated less frequently. When $D$ becomes too weak (i.e. fails to distinguish between real and fake inputs), then the opposite happens.
We found this new modification to be critical for ensuring that the $D$ does not overwhelm $G$ 
during training.

We also use the traditional adaptive discriminator augmentation \cite{ada_karras2020training} 
where we apply the following transforms with the same probability $p$:
(i) adding Gaussian noise with $\sigma = 0.05$, (ii) random scaling by a factor of $1 \pm 0.05$, and (iii) randomly replacing a subsequence of frames from the generated speech features with a subsequence of frames taken from a real speech feature sequence.
This last augmentation is based on the fake-as-real GAN method \cite{fake-as-real_tao2020alleviation} and is important to prevent gradient explosion later in training.

For the anti-aliasing LPF filters we use
windowed \texttt{sinc} filters with a width-9 Kaiser window \cite{kaiser1966digital}.
For the generator, the first \texttt{Style Block} has a cutoff 
at
$f_c = 0.125\ \text{cycles/sample}$ which is increased in an even logarithmic scale to $f_c=0.45\ \text{cycles/sample}$ in the second-to-last layer, keeping this value for the last two layers to fill in the last high frequency detail.
Even in these last layers we use a cutoff below the Nyquist frequency.
For the discriminator we are less concerned about aliasing as 
it does not generate a sequence, so we
use a 
$f_c=0.5\ \text{cycles/sample}$ 
cutoff
for all \texttt{ConvD Block}s.

All models are trained on a single NVIDIA Quadro RTX 6000 using PyTorch 1.11.
% \maybeshorten{Trained models and code are available at \url{https://slt2022blind.github.io/slt2022/}.}

%% file: 5_results.tex
\setlength{\tabcolsep}{3.2pt}
\begin{table}[!b]
    \renewcommand{\arraystretch}{1.2}
    \centering
    \caption{
        Results measuring the quality and diversity of generated samples from unconditional speech synthesis models together with train/test set toplines for the SC09 dataset. Subjective MOS values with 95\% confidence intervals are shown.
    }
    \tablecaptionsep
    \eightpt
    \label{tab:1_quality_diversity}
    
    \begin{tabularx}{1.0\linewidth}{@{}
        L
        S[table-format=1.2]
        S[table-format=3.1]
        S[table-format=1.2]
        S[table-format=1.2]
        S[table-format=1.2]
        S[table-format=1.2(2),
            table-figures-uncertainty=1,
            separate-uncertainty = true]
        @{}}
    \toprule
    Model & {IS$\ \uparrow$} & {mIS $\ \uparrow$} & {FID$\ \downarrow$} & {AM$\ \downarrow$} & {eMOS$\ \uparrow$} & {MOS$\ \uparrow$} \\
    \midrule
    \textit{Train set} & 9.37 & 237.6 & 0 & 0.20 & 2.41 & 3.74(12)\\
    \textit{Test set} & 9.36 & 242.3 & 0.01 & 0.20 & 2.43 & 3.88(12)\\
    \midrule
    WaveGAN \cite{adv_audio_synth_donahue2018adversarial} & 4.45 & 34.6 & 1.77 & 0.81 & 1.06 & 2.88(16) \\
    DiffWave \cite{diffwave_kong2020} & 5.13 & 49.6 & 1.68 & 0.68 & 1.66 & 3.43(14) \\
    SaShiMi \cite{sashimi_goel2022s} & 3.74 & 18.9 & 2.11 & 0.99  & 1.58 & 3.19(15) \\
    SaShiMi+DiffWave & 5.44 & 60.8 & 1.01 & 0.61 & 1.89 & 3.33(12)\\
    \modelname{} (mel-spec.) & 7.02 & 162.8 & 0.56 & 0.36 & 1.76 & 3.51(13)\\
    \modelname{} (HuBERT) & \ubold 7.67 & \ubold 226.7 & \ubold 0.14 & \ubold 0.26 & \ubold 1.99 & \ubold 3.68(13) \\
    \bottomrule
    \end{tabularx}
    % \aftertableskip
\end{table}

\setlength{\tabcolsep}{6pt}

\begin{table}[!b]
    \renewcommand{\arraystretch}{1.2}
    \centering
    \caption{
        Latent-space disentanglement and speed metrics. Speed is measured as the number of samples that can be generated per unit time on a single NVIDIA Quadro RTX 6000 using a batch size of 1 given in ksamples/sec.
        Some models do not have a $W$-space (WaveGAN) or any continuous latent space (SaShiMi).} 
    \tablecaptionsep
    \label{tab:2_disentanglement}
	\eightpt
	\begin{tabularx}{0.99\linewidth}{@{}Lccccr@{}}
		\toprule
		& \multicolumn{2}{c}{{Path length $\downarrow$}} & \multicolumn{2}{c}{Separability $\downarrow$} &  \\
		\cmidrule(l){2-3}  \cmidrule(l){4-5}
		Model & {$Z$} & {$W$} & {$Z$} & {$W$} & {Speed $\uparrow$}  \\
		\midrule
WaveGAN \cite{adv_audio_synth_donahue2018adversarial} & \ubold 1.03\hphantom{$\cdot \text{10}^{\text{0}}$}  & {---} & 4.86 & {---} & \textbf{2214.71} \\
DiffWave \cite{diffwave_kong2020} & 2.72$\cdot \text{10}^{\text{6}}$ & 7.27$\cdot \text{10}^{\text{5}}$  & 6.09 & 6.58 & 0.83 \\
SaShiMi \cite{sashimi_goel2022s} & {---} & {---} & {---} & {---} & 0.14  \\
SaShiMi+DiffWave & 2.89$\cdot \text{10}^{\text{6}}$ & 1.24$\cdot \text{10}^{\text{6}}$ & 4.07 & 2.34 & 0.47 \\
\modelname{} (mel-spec.) & 6.77$\cdot \text{10}^{\hphantom{\text{1}}}$ & 3.21$\cdot \text{10}^{\hphantom{\text{1}}}$ & 1.81 & 1.01 & 875.45 \\
\modelname{} (HuBERT) & 3.50$\cdot \text{10}^{\hphantom{\text{1}}}$ & \ubold 1.84$\cdot \text{10}^{\hphantom{\text{1}}}$ & \textbf{1.40} & \textbf{1.00} & 816.27 \\
		\bottomrule
	\end{tabularx}
\end{table}

\begin{figure*}[t!]
\centering
     \begin{subfigure}{0.85\textwidth}
         \centering
         \includegraphics[width=\textwidth]{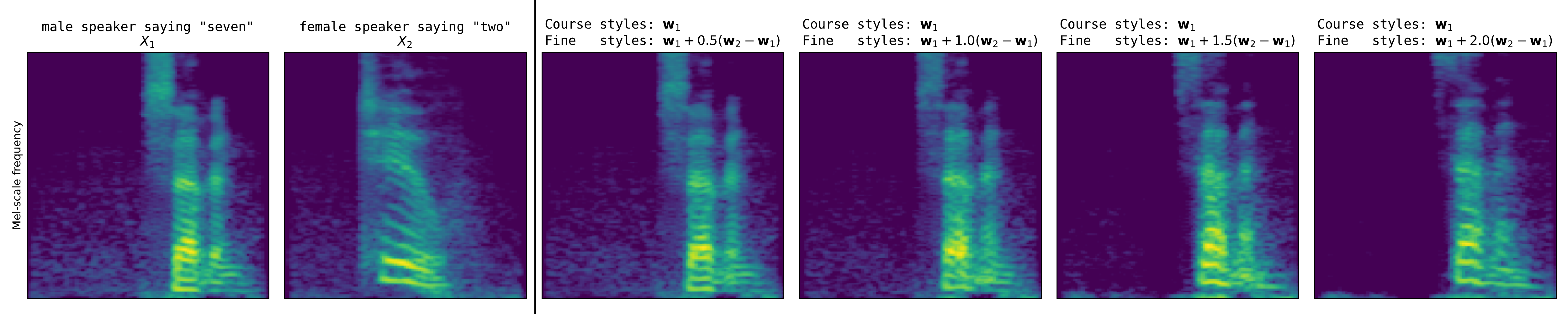}
         \caption{Voice conversion}
         \label{fig:6_vc}
     \end{subfigure}
     \begin{subfigure}{0.85\textwidth}
         \centering
         \includegraphics[width=\textwidth]{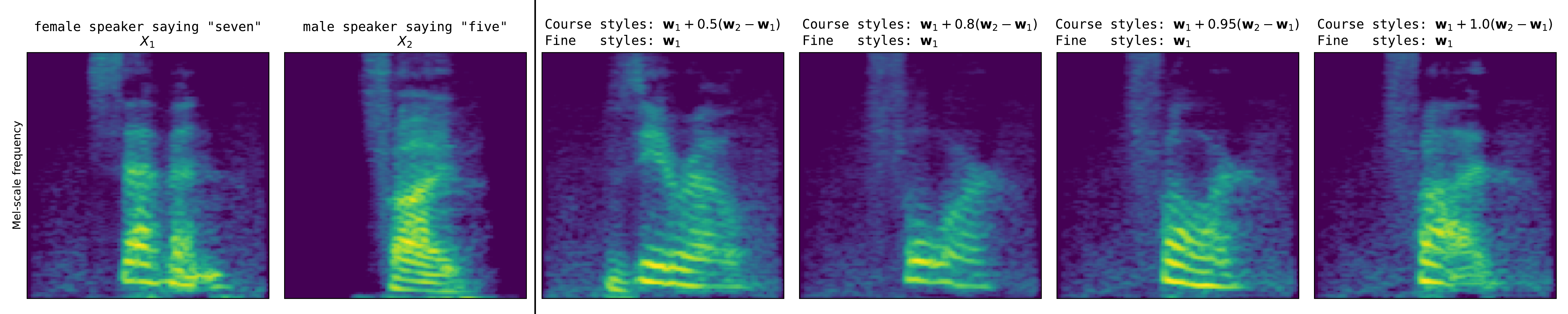}
         \caption{Speech editing: Digit/content conversion}
         \label{fig:6_spk_edit}
     \end{subfigure}
\caption{
    Examples of \modelname{} (HuBERT variant)
    performing unseen tasks on unseen speakers from the SC09 test set: (a) voice conversion and (b) speech editing.
    We encourage the reader to listen to audio samples at {\footnotesize \url{https://rf5.github.io/slt2022-asgan-demo}}.
}
\vspace*{-8pt}
\label{fig:6_unseen}
\end{figure*}

Table~\ref{tab:1_quality_diversity}
compares
previous state-of-the-art unconditional speech synthesis approaches to the newly proposed ASGAN.
As a reminder, IS, mIS, FID and AM measure generated speech diversity and quality relative to the test set; eMOS and MOS are measures of generated speech naturalness.
We see that both variants of \modelname{}
outperforms
the other models
on most metrics.
The HuBERT variant of \modelname{} in particular
performs best across
all metrics. 
The improvement of the HuBERT \modelname{} over the mel-spectrogram variant is likely
because the high-level HuBERT speech representations
make it easier for the model to disentangle common factors of speech variation. 
The previous best unconditional synthesis model, SaShiMi+DiffWave, still outperforms the other baseline models, 
and it appears to have comparable naturalness (similar eMOS and MOS) to the mel-spectrogram \modelname{} variant.
However, it appears to match the test set more poorly than either \modelname{} variant on 
the other diversity metrics.

Latent space disentanglement and generation speed of each model are measured
in Table~\ref{tab:2_disentanglement}.
These results are more mixed, with WaveGAN being the fastest model and the one with the shortest latent $Z$-space path length.
However, this is somewhat misleading since WaveGAN's samples have low quality
and poor diversity compared to the other models
(see Table~\ref{tab:1_quality_diversity}).
This means that WaveGAN's latent space 
is
a poor representation of the true distribution of speech in the SC09 dataset, allowing it to have a very small path length as most paths
do not span a diverse set of speech variation. 

In terms of linear separability,
\modelname{} again yields substantial improvements over existing models.
The results
confirm that 
\modelname{} has indeed learned
a disentangled latent space -- a primary motivation for the model's design.
Specifically,
this shows that the idea from image synthesis of
using the latent $\mathbf{w}$ vector to linearly modulate convolution kernels
can also be applied to speech.
This level of disentanglement allows \modelname{} to be applied to tasks unseen during training,
as described in the next 
section.

Regardless of performance, the speed of all the convolutional GAN models (WaveGan and \modelname{}) is significantly better than the diffusion and autoregressive models.
This highlights an additional 
benefit of utilizing convolutional GANs that produce utterances in a single inference call, as opposed to the many inference calls necessary 
with autoregressive or diffusion modelling.

%% file: 6_zeroshot.tex
To further showcase the disentangled latent space learned by \modelname{}, here we qualitatively consider how it can be used
to perform voice conversion and speech editing without any further
training.
Our goal is not to achieve state-of-the-art results on these tasks or to present a complete quantitative evaluation, but simply to illustrate
the \textit{ability} for our model to transfer to
these unseen tasks.

For
these
tasks we wish to modify an already existing utterance which has not been produced by the generator $G$.
To do this, we need to map the
speech features back to the $G$'s
latent $W$-space. 
This is done
using a method similar to \cite{stylegan2_karras2020analyzing} whereby we optimize a $\mathbf{w}$ vector while keeping $G$ and the
speech feature sequence $X$ fixed.
Concretely, $\mathbf{w}$ is initialized to the mean $\Bar{\mathbf{w}}= \mathbb{E}_\mathbf{z} [ W(\mathbf{z}) ]$
and then fed through the network to produce a candidate sequence $\Tilde{X}$.
An $L_2$ loss
between
the candidate sequence $\Tilde{X}$ and the target sequence $X$
is then optimized
using Adam
with the settings from \cite{stylegan2_karras2020analyzing}.

We can modify several aspects of speech from seen or unseen speakers by performing style mixing \cite{stylegan1_karras2019style}.
Concretely, given speech features for two utterances $X_1$ and $X_2$ from potentially unseen speakers,
we first project them to the latent space, obtaining $\mathbf{w}_1$ and $\mathbf{w}_2$.
We can then use different $\mathbf{w}$ vectors as the input into each \texttt{Style Block} in Fig.~\ref{fig:1_model_arch}. 
According to our design 
motivation in Sec.~\ref{sec:3_conv_encoder}, the \textit{course styles} (e.g. which word is said) are captured
in the earlier layers and the \textit{fine styles} (e.g. speaker identity, tone)
in later layers.
So, we can perform voice conversion from $X_1$'s speaker to $X_2$'s speaker by simply replacing the $\mathbf{w}$ vector
in
the last 5 modulated convolutions (fine styles) with $\mathbf{w}_2$, while using $\mathbf{w}_1$ in the earlier blocks {(course styles)}.
By doing the opposite, we can also do
speech editing -- the task of replacing the content of the words spoken (replacing $\mathbf{w}$ for course styles), but leaving the speaker identity intact (retaining $\mathbf{w}$ for fine styles).
Furthermore, because the $W$-space is continuous, we can 
interpolate between replacing the course and fine styles to achieve varying degrees of
voice conversion or speech editing.

An example of these tasks
on unseen speakers on the SC09 test set is shown in Fig.~\ref{fig:6_unseen}.
For these examples we use the truncation trick \cite{stylegan1_karras2019style} in the $W$-space with truncation $\psi = 0.3$.
We encourage the reader to listen to the samples (link given in the caption).

%% file: 7_conclusion.tex
We introduced \modelname{}, a model for unconditional speech synthesis designed to learn a disentangled latent space.
Specifically, we adapted existing and incorporated new GAN design and training techniques to enable \modelname{} to outperform existing
autoregressive and diffusion
models. 
Experiments
on the SC09 dataset validated this design, demonstrating that \modelname{} 
outperforms previous
state-of-the-art
models
on most metrics,
while also being substantially faster.
Further experiments also demonstrated the benefit of the disentangled latent space -- \modelname{} can, without any additional training, perform voice conversion and speech editing
in a zero-shot fashion
through linear operations in its latent space.

One major limitation of our work is
scale:
once trained, \modelname{} can only generate utterances of a fixed length, and the model struggles to generate coherent full sentences on datasets with
longer utterances
(a limitation shared by existing unconditional synthesis models).
Future work will aim to address this shortcoming by considering which aspects of \modelname{} can be simplified or removed to improve scaling.
Future work will also perform more thorough subjective evaluations to quantify how \modelname{} performs on unseen tasks.